\documentclass{article}
\usepackage{spconf,amsmath,graphicx,hyperref}
\usepackage{siunitx}
\usepackage{siunitx}
\usepackage{booktabs}
\usepackage{pgfplots}
\pgfplotsset{compat=1.18}
\newcommand{\datasetname}{AI-OpenBMAT}

\usepackage{booktabs}
\usepackage{tabularx}   % optional, for automatic column widths
\usepackage{siunitx}    % optional, nice numeric alignment
\usepackage{stfloats}
\usepackage{placeins}   % provides \FloatBarrier
\usepackage{soul}
\sethlcolor{BurntOrange}
\usepackage[usenames,dvipsnames]{xcolor}

\title{AI-generated music detection in broadcast monitoring}
\name{
David López-Ayala$^{1}$,
Asier Cabello$^{2}$,
Pablo Zinemanas$^{2}$,
Emilio Molina$^{2}$,
Mart\'in Rocamora$^{1}$
}
\address{
    $^1$ Music Technology Group, Universitat Pompeu Fabra, Barcelona, Spain\\
    $^2$ BMAT Licensing S.L., Barcelona, Spain\\
    \tt\normalsize\{jorgedavid.lopez,martin.rocamora\}@upf.edu \{acabello,pzinemanas,emolina\}@bmat.com
}

\begin{document}
%\ninept
%
\maketitle
\def\abstract{\begin{center}
{\bf ABSTRACT\vspace{-.5em}\vspace{0pt}}
\end{center}}
\def\endabstract{\par}
\begin{abstract}

AI music generators have advanced to the point where their outputs are often indistinguishable from human compositions. While detection methods have emerged, they are typically designed and validated in music streaming contexts with clean, full-length tracks. Broadcast audio, however, poses a different challenge: music appears as short excerpts, often masked by dominant speech, conditions under which existing detectors fail. In this work, we introduce \datasetname{} \footnote{github.com/DaveLoay/AI-OpenBMAT.}, the first dataset tailored to AI-generated music detection in a broadcast setting. 
It contains 3,294 one-minute audio excerpts (54.9 hours) that follow the duration patterns and loudness relations of real television audio, combining human-made production music with stylistically matched continuations generated with Suno v3.5. We benchmark a CNN baseline and state-of-the-art SpectTTTra models to assess SNR and duration robustness, 
% through SNR sweeps, temporal granularity tests, 
and evaluate on a full broadcast scenario. Across all settings, models that excel in streaming scenarios suffer substantial degradation, with F1-scores dropping below 60\% when music is in the background or has a short duration. These results highlight speech masking and short music length as critical open challenges for AI music detection, and position \datasetname{} as a benchmark for developing detectors capable of meeting industrial broadcast requirements.

\end{abstract}
\section{Introduction}

The rapid advances in generative models have enabled the creation of music that is difficult to distinguish from human compositions~\cite{Agostinelli2023, Evans2024, caillon2021ravevariationalautoencoderfast, song2020generativemodelingestimatinggradients}. While this opens creative opportunities, it also creates operational and legal challenges for stakeholders such as streaming services, broadcasters, and rights holders, who require reliable detectors of AI-generated music in the real world \cite{GEMA2025_AIlawsuit}. However, most existing datasets and detection models have been developed with a streaming-centric view of the problem—i.e., inputs whose durations are comparable to an average song (roughly 3--4 minutes) and that present music as a prominent, foreground signal. 
This is driven by the growing prevalence of AI-generated music, as reports estimate that nearly 30\% of music uploads to platforms such as Deezer are AI-generated~\cite{MusicAlly2025DeezerAI, Ingham2025NearlyThird}.

However, broadcasting differs substantially from this streaming scenario: music typically appears as short excerpts —often only a few seconds— instead of full-length tracks; and music is frequently in the background beneath dominant speech or sound effects, producing low signal-to-noise ratios and complex mixtures. Motivated by the practical need for reliable AI-generated music detection in broadcast monitoring, we observed that detectors trained under streaming assumptions degrade markedly when deployed in broadcast settings. In this work, we rigorously demonstrate these limitations and provide a benchmark dataset for developing more robust detection methods for broadcast monitoring.

Prior work on AI-music detection showed that shallow CNN classifiers can attain very high accuracy in controlled settings, but exhibit limited robustness to simple audio transformations and often generalize poorly to unseen generative models~\cite{afchar2025aigeneratedmusicdetectionchallenges}. These classifiers seem to exploit artifacts unintentionally left by the decoder architectures of the generative models, % (through upsampling and stransposed-convolution stages), 
that act as model-dependent cues for detection~\cite{afchar2025fourierexplanationaimusicartifacts}. A more sophisticated approach for detection was proposed in~\cite{rahman2025sonicssyntheticidentifying}, the SpectTTTra architecture, that exploits long-range musical context via spectro-temporal tokenization of audio, and improves over conventional CNN and Transformer-based models. The work also introduced the SONICS dataset, comprising real and AI-generated songs from Suno and Udio platforms, totaling approximately 4,751 hours of audio. Another large dataset of over 1,770 hours for the detection of AI‑generated music was compiled in~\cite{Vila-2025}, of which one third is human‑made, and two thirds comes from popular AI music platforms. The paper tests different AI music detectors (including a commercial baseline) and shows their vulnerability to various audio transformations. Some other resources that include AI-generated music, e.g., Sunocaps~\cite{CIVIT2024110743} and FakemusicCaps~\cite{comanducci2024fakemusiccapsdatasetdetectionattribution}, address attribution and text–to–music alignment, respectively, focusing on captioning and alignment rather than binary discrimination between human and AI audio. Yet, despite these efforts, the AI music-detection task in broadcast conditions remains largely unexplored.

To fill this gap, we build upon the OpenBMAT dataset for music detection~\cite{Meléndez-Catalán-2019}, which contains over 27 hours of TV broadcast audio in one-minute-long expert segments and includes annotations on the loudness of music relative to other simultaneous non-music sounds. In this work, we introduce \datasetname{}, a dataset constructed to evaluate AI-generated music detection under broadcast conditions. \datasetname{} contains 54.9 hours of audio, as one-minute excerpts, created by pairing 1,647 human-made music production tracks with stylistically matched continuations generated by Suno v3.5, then remixing these pairs with speech audio using OpenBMAT annotations on segment structures and realistic loudness statistics~\cite{Meléndez-Catalán-2019}. Using \datasetname{}, we systematically benchmark state-of-the-art detectors (SpectTTTra variants and a CNN baseline) across three controlled experiments: (i) simulate speech masking at different SNRs, (ii) measure sensitivity to short-duration input, and (iii) evaluate models on full broadcast conditions. We reveal substantial performance drops and highlight the need for improved detection models suited to broadcast monitoring.

\section{DATASET}
%\section{\datasetname}
\label{sec:method}

\subsection{\datasetname{} overview}

\datasetname{} is explicitly designed to mimic the statistical structure of broadcast music in OpenBMAT. To achieve this, we first analysed OpenBMAT to extract the temporal patterns and the music–speech loudness / SNR distributions that characterise TV audio \cite{Meléndez-Catalán-2019}. These statistics define the parametrisation used to generate each simulated broadcast audio segment. Specifically, each \datasetname{} example is a one-minute broadcast-like audio segment constructed by mixing a production music track with speech-only audio according to the temporal pattern and target SNR values derived from OpenBMAT. The production music sources are 476 human-composed production tracks from Epidemic Sound used in the BAF dataset \cite{cortes2022BAF}, which is the first public dataset for music monitoring in broadcast, thus closely matching the style of music used in television. 

For every human-composed track, we create a corresponding AI-generated version by using Suno v3.5's \textit{\textbf{extend}} feature to continue the original audio. Generating continuations from the human track ensures high stylistic, timbral, and semantic correspondence among pairs, minimising confounds arising from genre or instrumentation differences between human-made and AI-generated music. %and isolating the detection task to artifacts introduced by the generative process itself. 
The final dataset contains 1,647 corresponding human/AI pairs (one-minute each), totaling 54.9 hours of mixed, broadcast-like audio. These design choices make \datasetname{} suitable for evaluating detector robustness to the temporal duration of the audio input and adverse acoustic masking in a controlled, repeatable manner.

\subsection{Dataset creation process}
\label{sec:methodology}
We build each \datasetname{} excerpt by reusing the segment annotations provided by OpenBMAT. OpenBMAT annotations consist of a non-overlapping segmentation that covers the full audio waveform and assigns one of six discrete classes that describe the \emph{relative} loudness of music. The original labeling taxonomy used is: \texttt{no-music}, \texttt{music}, \texttt{background music (bg)}, \texttt{similar}, \texttt{foreground music (fg)}, and \texttt{low background music (bgvl)}. Annotators followed a minimum-segment-length rule and a small set of merging heuristics to avoid an excessive number of very short changes: groups of non-music shorter than 1\,s may be merged when they are closer than 1\,s or have similar loudness, and remaining unlabeled regions are finally marked as \texttt{music}. From each OpenBMAT reference excerpt, we extract the segment boundaries, the taxonomy class, and the reference segment energy/loudness statistics. This information determines the number of segments, the segment durations, and the target loudness/SNR relations that we reproduce when constructing \datasetname{} audio files.

Each one-minute \datasetname{} file is assembled by copying the segment structure of a reference OpenBMAT excerpt (number of segments, segment types, and segment durations) and synthesizing segments that replicate the reference's measured loudness and SNRs. The synthesised segments are then concatenated to produce a one-minute audio file. Human music clips were drawn from the Epidemic Sound selection within the BAF dataset~\cite{cortes2022BAF}. 
For AI music, we generated a single continuation per human track using Suno v3.5's extend mode so that the generated audio continues the original production music. 
Non-music audio is taken from OpenBMAT from segments labeled as \texttt{no-music}, and almost always contains speech, with a few exceptions of other sounds, such as hand clapping or environmental sounds.

Segment construction proceeds as follows. For segments that require both music and speech, we first select candidate music and speech fragments, unit-normalize their levels, and then set the music–speech mix ratio to a target SNR (music as signal, speech as noise). Target SNRs were chosen per taxonomy class as: \texttt{foreground music} $-5\,$dB, \texttt{similar} $0\,$dB, and \texttt{low background music} $-10\,$dB. The actual SNR of the segment is randomly sampled from a uniform distribution with the target SNR as the mean value and a \(\pm 3\,\)dB range. Purely \texttt{music} or \texttt{no-music} segments contain a single source, so no SNR is involved. After mixing, each segment is RMS-matched to the reference segment energy measured from the OpenBMAT excerpt. 

To avoid audible discontinuities at segment boundaries when concatenating segments, we apply 20\,ms linear fades at segment in/outs and a 50\,ms crossfade between consecutive segments. Final tracks are trimmed to exactly 60.0\,s. All mixes are exported as WAV audio (mono, 22.05\,kHz, 16-bit).

\subsection{Segment distribution} %Analyzing \datasetname{}}
\begin{figure}[t]
\centering
\begin{tikzpicture}
\begin{axis}[
    width=0.45\textwidth,
    height=0.25\textwidth,
    ybar,
    bar width=14pt,
    enlarge x limits=0.15,
    ylabel={Total duration (\%)},
    ymin=0, ymax=55,
    ytick={0,10,20,30,40,50},
    axis x line*=bottom,
    axis y line*=left,
    grid=major,
    grid style={line width=.08pt, draw=gray!30},
    symbolic x coords={bgvl-music,fg-music,music,similar,bg-music,no-music},
    xtick=data,
    xticklabel style={rotate=20, anchor=north east, font=\small},
    every axis plot/.append style={draw=black,fill=gray!30},
    nodes near coords,
    nodes near coords style={font=\footnotesize, /pgf/number format/fixed, /pgf/number format/precision=1},
    tick align=outside
]

\addplot coordinates {
  (bgvl-music,4.441741)
  (fg-music,5.260267)
  (music,10.689016)
  (similar,11.165133)
  (bg-music,20.1)
  (no-music,48.418614)
};
\end{axis}
\end{tikzpicture}
\caption{Distribution of \datasetname{} by segment class. } % (percentage of total duration). \texttt{bgvl-music} denotes OpenBMAT's \emph{Low Background Music} class.}
\label{fig:ds_distribution}
\end{figure}
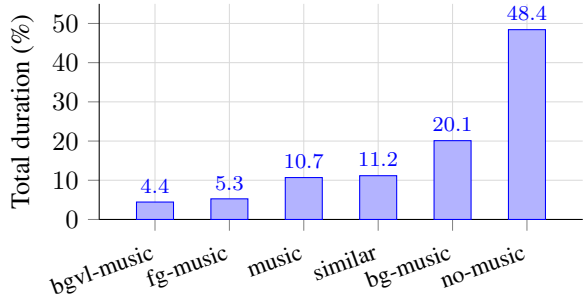

\datasetname{} is composed of 18,848 annotated segments across its 3,294 tracks. The distribution across segment class is shown in Fig.~\ref{fig:ds_distribution}. Mean segment durations per class are: \texttt{no-music} 37.0\,s, \texttt{music} 8.5\,s, \texttt{background music} 7.6\,s, \texttt{low background music} 6.6\,s, \texttt{similar} 5.9\,s, and \texttt{foreground music} 4.2\,s. Although the dataset provides a broad range of SNR conditions, it is dominated by short speech (i.e.,\ non-music) segments, which is consistent with a typical broadcast scenario.

\section{Experiments}
\label{sec:experiments}

The central hypothesis of this work is that current AI-music detectors, developed and validated in a streaming-centric scenario, degrade under broadcast conditions characterised by (i) masking speech, where music is frequently masked by dominant foreground speech, yielding low SNR; and (ii) short-duration music, where music appears as short excerpts rather than full-length tracks. To test this hypothesis, we design three experiments: SNR robustness, duration robustness, and the full broadcast condition. The first two experiments seek to isolate the effect of masking and duration independently. The latter experiment uses the proposed \datasetname{} for simulating the broadcast scenario.

\subsection{AI-music detection models}

We benchmark state-of-the-art SpectTTTra~\cite{rahman2025sonicssyntheticidentifying} pre-trained models in the 5s configuration and a shallow CNN baseline. The SpectTTTra models were trained on the SONICS dataset, and can be distinguished by the size of their spectral ($f$) and temporal ($t$) patches: SpectTTTra-$\alpha$ ($f=1, t=3$), SpectTTTra-$\beta$ ($f=3, t=5$), and SpectTTTra-$\gamma$ ($f=5, t=7$).  The CNN is a six-layer 2D convolutional network with 64 filters in each layer, kernel size $3$, and finishes with two fully connected layers. It was trained on a private dataset comprising 27h of human and 27h of AI (Suno v3.5) music, achieving $99.97\%$ accuracy on its held-out test set.

\subsection{SNR robustness}
% \paragraph*{SNR robustness} \hskip -0.5em 
The first experiment seeks to evaluate robustness to speech interference. To do that, we constructed 5-second mixtures of music and broadcast speech at controlled SNR values ranging from $+\infty$ (music alone) down to $-30$\,dB in 5\,dB steps, with headroom of $-3$\,dB to avoid clipping. As for music, we utilized human music from Epidemic Sound and AI-generated continuations from Suno v3.5. Speech audio was taken from OpenBMAT's \texttt{non-music} segments. Loudness was normalized (LUFS) prior to mixing, and per-second SNR was enforced within $\pm 1$\,dB of the target. The detection models were tested on a dataset of 9,554 audio mixtures at different SNRs (\SI{796}{min} total), comprising 4,777 Human/AI-generated music track pairs.

\subsection{Duration robustness}
The second experiment assesses the impact of input duration. While all detector models operate on fixed 5\,s windows, music segments in a broadcast setting may be shorter.  
We built a dataset of 952 tracks, comprising 476 pairs of human music tracks from Epidemic Sound and AI-generated continuations from Suno v3.5. For each track in the dataset, we tested the detection models with windows containing 5, 4, 3, 2, 1, 0.5, and 0.2 seconds of the actual audio track, and zero-padding the remaining portion of the window to its total length.

\subsection{Broadcast scenario}
Finally, we evaluate the detection models on the full \datasetname{} dataset of 1,647 human--AI pairs (\SI{54.9}{h}, 3,294 tracks in total). Each complete one-minute excerpt of the dataset is classified using the detection models, while accounting for variable segment lengths and realistic SNR distributions. This experiment represents the most ecologically valid setting, directly mirroring broadcast conditions.

F1-scores are computed time-wise, i.e., they measure the proportion of total audio duration correctly classified as human or AI-generated. Inference is performed with a fixed 5\,s analysis window and a hop size of 1\,s: each 1\,s region is covered by multiple overlapping windows, and its final prediction is obtained as the moving average of the predictions from all windows that include that second. The predicted label for each second is then compared with the ground-truth annotation to compute F1 Scores. We also experimented with hop sizes ranging from 0.5\,s to 5\,s, and although shorter hops yield slightly better results, the improvements are marginal; thus, we chose to report only the 1\,s case. % for consistency.

\section{Results and discussion}
Across all experiments, we report F1-score, which balances precision and recall and is more informative than accuracy under class-imbalanced conditions.

\subsection{SNR robustness}
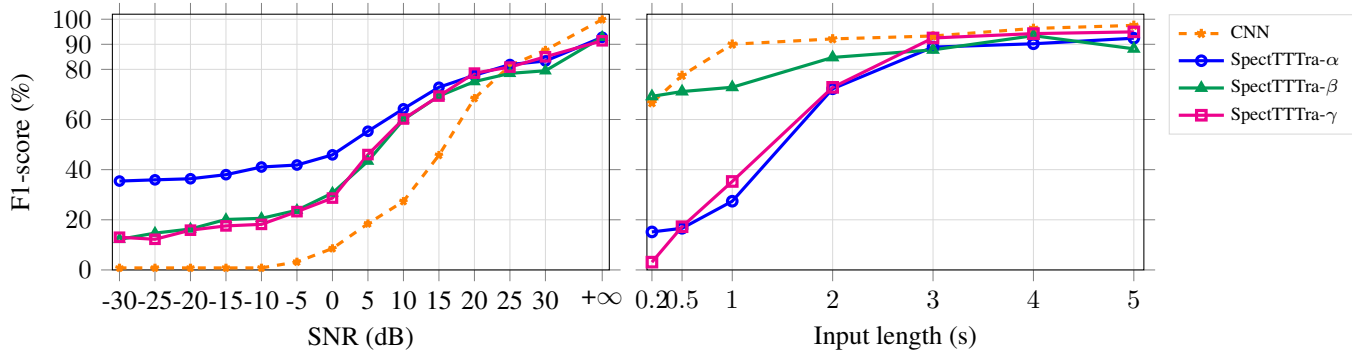
\begin{figure*}[t]
\centering
\begin{tikzpicture}

    % 1. Define a consistent style for all models (Color, Line style, Marker)
    % Order: CNN, Alpha, Beta, Gamma
    \pgfplotscreateplotcyclelist{model_styles}{
        orange, dashed, mark=star, very thick\\          % CNN
        blue, solid, mark=o, very thick\\               % Alpha
        ForestGreen, solid, mark=triangle, very thick\\         % Beta
        magenta, solid, mark=square, very thick\\ % Gamma
    }

    % --- CHART 1: SNR (Left) ---
    \begin{axis}[
        name=plot1,
        width=0.46\textwidth,
        height=0.28\textwidth,
        xlabel={SNR (dB)},
        ylabel={F1-score (\%)},
        xmin=-31, xmax=39,
        ymin=0, ymax=102,
        xtick={-30,-25,-20,-15,-10,-5,0,5,10,15,20,25,30,38},
        xticklabels={
            -30,-25,-20,-15,-10,-5,0,5,10,15,20,25,30,+$\infty$
        },
        ytick={0,20,40,60,80,90,100},
        grid=both,
        grid style={line width=.1pt, draw=gray!20},
        major grid style={line width=.2pt,draw=gray!30},
        tick align=outside,
        cycle list name=model_styles,
        mark options={scale=0.8}
    ]

    % CNN
    \addplot coordinates {
     (38.0,99.89) (30.0,87.671) (25.0,81.446) (20.0,68.449) (15.0,45.768)
     (10.0,27.368) (5.0,18.45) (0.0,8.56)
     (-5.0,3.2) (-10.0,0.81) (-15.0,0.806) (-20.0,0.806)
     (-25.0,0.803) (-30.0,0.803)
    };

    % Alpha
    \addplot coordinates {
     (38.0,92.8) (30.0,83.333) (25.0,81.922) (20.0,77.619) (15.0,72.861)
     (10.0,64.267) (5.0,55.319) (0.0,45.902)
     (-5.0,41.848) (-10.0,41.053) (-15.0,37.995) (-20.0,36.364)
     (-25.0,35.938) (-30.0,35.45)
    };

    % Beta
    \addplot coordinates {
     (38.0,92.8) (30.0,79.439) (25.0,78.404) (20.0,75.122) (15.0,69.347)
     (10.0,59.947) (5.0,43.402) (0.0,30.671)
     (-5.0,23.794) (-10.0,20.579) (-15.0,20.13) (-20.0,16.327)
     (-25.0,14.685) (-30.0,12.186)
    };

    % Gamma
    \addplot coordinates {
     (38.0,91.4) (30.0,85.022) (25.0,80.909) (20.0,78.505) (15.0,69.347)
     (10.0,60.215) (5.0,46.018) (0.0,28.664)
     (-5.0,23.256) (-10.0,18.182) (-15.0,17.568) (-20.0,15.862)
     (-25.0,12.23) (-30.0,13.043)
    };

    \end{axis}

    % --- CHART 2: LENGTH (Right) ---
    \begin{axis}[
        name=plot2,
        at={(plot1.south east)},
        xshift=0.5cm,
        width=0.46\textwidth,
        height=0.28\textwidth,
        xlabel={Input length (s)},
        xmin=0.15, xmax=5.1,
        ymin=0, ymax=102,
        xtick={0.2,0.5,1,2,3,4,5},
        ytick={0,20,40,60,80,90,100},
        yticklabels={,,},
        grid=both,
        grid style={line width=.1pt, draw=gray!20},
        major grid style={line width=.2pt,draw=gray!30},
        tick align=outside,
        cycle list name=model_styles,
        mark options={scale=0.9},
        legend style={
            at={(1.05,1)},
            anchor=north west,
            font=\scriptsize,
            cells={anchor=west},
            draw=gray!30
        }
    ]

    % CNN
    \addplot coordinates {
     (5.0,97.54) (4.0,96.32) (3.0,93.31) (2.0,92.15)
     (1.0,90.02) (0.5,77.57) (0.2,66.66)
    };
    \addlegendentry{CNN}

    % Alpha
    \addplot coordinates {
     (5.0,92.43) (4.0,90.21) (3.0,88.88) (2.0,72.26)
     (1.0,27.41) (0.5,16.61) (0.2,15.16)
    };
    \addlegendentry{SpectTTTra-$\alpha$}

    % Beta
    \addplot coordinates {
     (5.0,88.20) (4.0,93.44) (3.0,87.78) (2.0,84.76)
     (1.0,72.86) (0.5,71.15) (0.2,69.24)
    };
    \addlegendentry{SpectTTTra-$\beta$}

    % Gamma
    \addplot coordinates {
     (5.0,94.95) (4.0,94.26) (3.0,92.52) (2.0,72.83)
     (1.0,35.29) (0.5,17.24) (0.2,3.10)
    };
    \addlegendentry{SpectTTTra-$\gamma$}

    \end{axis}

\end{tikzpicture}
\caption{Impact of environmental factors on performance. \textbf{Left:} F1-score versus SNR. \textbf{Right:} F1-score versus input audio duration. The shared legend applies to both figures.}
\label{fig:combined_analysis}
\end{figure*}

All models exhibit notable performance degradation as SNR decreases (Fig.~\ref{fig:combined_analysis}). At clean conditions ($+\infty$\,dB), scores are high, with the CNN baseline achieving 99.97\% and SpectTTTra models nearly 93\%. However, even at $+30$\,dB, where the music is dominant, and speech is almost imperceptible, all models suffer about a 10\% drop. As SNR decays further, CNN performance deteriorates most steeply. Among the SpectTTTra models, the $\alpha$ variant is most robust, decaying more slowly than the $\beta$ and $\gamma$ models.

\subsection{Duration robustness}
Reducing audio input duration strongly impacts detection (Fig.~\ref{fig:combined_analysis}). Interestingly, the CNN remains relatively stable down to 1s, with a performance drop of less than 10\%, but falls sharply below 1s, reaching 66\% at 0.2s. SpectTTTra models are more sensitive: both $\alpha$ and $\gamma$ drop to 72\% at 2s and deteriorate further for shorter lengths. This indicates that SpectTTTra models rely more heavily on long-range context, whereas the CNN retains discriminative ability for shorter excerpts, highlighting complementary weaknesses.

\subsection{Broadcast scenario}

When evaluated on the full \datasetname{} dataset, performance drops considerably compared to the typical streaming scenario (Table ~\ref{tab:full_results}). 

Under broadcast monitoring conditions, the best SpectTTTra model achieves an overall F1-score of only 61.1\%, while the CNN baseline falls dramatically to 27.6\%. The per-class breakdown aligns with the trends observed in the SNR and duration experiments. For instance, the SpectTTTra models' performance remains high for \texttt{music} and \texttt{foreground music (fg)} classes (above 78\%), where music is clean and dominant; drops by nearly 30 percentage points for the \texttt{similar} class, where speech and music are at comparable levels; and deteriorates drastically for \texttt{background (bg)} and \texttt{low background (bgvl)}, confirming that background speech-masked music is very challenging. The CNN baseline shows similar behaviour, though the performance drop is even more noticeable.

\begin{table}[t]
\centering
\small
\setlength{\tabcolsep}{6pt}
\resizebox{0.48\textwidth}{!}{%
\begin{tabular}{@{} l c @{\hspace{10pt}} c c c c c @{}}
\toprule
\textbf{Model} & \textbf{Overall} & \multicolumn{5}{c}{\textbf{Per-class F1 }} \\
\cmidrule(lr){3-7}
 & \textbf{F1} & \textbf{bg} & \textbf{bgvl} & \textbf{fg} & \textbf{music} & \textbf{similar} \\
\midrule
SpectTTTra-$\alpha$ & 57.6 & \textbf{54.3} & \textbf{47.0} & \textbf{84.4} & 88.5 & \textbf{61.7} \\
SpectTTTra-$\beta$  & 54.3 & 44.2 & 36.4 & 78.0 & 83.9 & 50.3 \\
SpectTTTra-$\gamma$ & \textbf{61.1} & 46.9 & 33.2 & \textbf{84.4} & \textbf{88.9} & 55.8 \\
\midrule
CNN                 & 27.6 & 13.4 & 3 & 33 & 63.1 & 13.6 \\
\bottomrule
\end{tabular}%
}
\caption{Overall and per-class F1-scores (\%) for all the AI-music detection models evaluated on \datasetname{}.}
\label{tab:full_results}
\end{table}

\section{Conclusions}
\label{sec:conclusions}

In this paper, we introduce \datasetname{}, the first dataset for end-to-end AI-music detection designed specifically for the broadcast monitoring scenario. By leveraging the structure and taxonomy of OpenBMAT and pairing human production tracks with stylistically matched continuations from Suno v3.5, the dataset provides a controlled yet ecologically valid benchmark that incorporates both short audio durations and adverse masking conditions.  
Through systematic experiments, we demonstrate that models that achieve near-perfect accuracy in streaming-centric scenarios, including a simple CNN baseline and the state-of-the-art SpectTTTra models, experience a substantial drop in performance under broadcast-like conditions. 
These findings confirm that broadcast monitoring requirements, e.g.,\ robustness to speech masking and short audio inputs, remain an open challenge. 
\datasetname{} therefore fills a critical gap by offering a benchmark that better reflects some industrially relevant use cases. 
We expect it to serve both as a % diagnostic 
tool to reveal the limitations %  failure modes
of existing detectors and as a foundation for developing methods that are robust to broadcast conditions.

\section{ACKNOWLEDGMENTS}
\label{sec:ACKNOWLEDGMENTS}
This work has been supported by the project "IA y Música: Cátedra en Inteligencia Artificial y Música (TSI-100929-2023-1)",
funded by the "Secretaría de Estado de Digitalización e Inteligencia Artificial and the Unión Europea-Next Generation EU".
% \vfill
% \pagebreak

\bibliographystyle{IEEEbib}
\bibliography{strings,refs}

\end{document}